\documentclass[twocolumn,showpacs,preprintnumbers,amsmath,amssymb,prb]{revtex4}
\usepackage{graphicx}% Include figure files
\usepackage{dcolumn}% Align table columns on decimal point
\usepackage{bm}% bold math

\begin{document}

\title{Nonequilibrium critical dynamics of the three-dimensional gauge glass}

\author{Federico Rom\'a}
\affiliation{Centro At{\'{o}}mico Bariloche, R8402AGP San Carlos
de Bariloche, R\'{\i}o Negro, Argentina}
\author{Daniel Dom\'{\i}nguez}
\affiliation{Centro At{\'{o}}mico Bariloche, R8402AGP San Carlos
de Bariloche, R\'{\i}o Negro, Argentina}

\date{\today}

\begin{abstract}

We study the non-equilibrium aging behavior of the gauge glass
model in three dimensions at the critical temperature. We perform
Monte Carlo simulations with a Metropolis update, and correlation
and response functions are calculated for different waiting times.
We obtain a multiplicative aging scaling of the correlation and
response functions,  calculating the aging exponent $b$ and the
nonequilibrium autocorrelation decay exponent $\lambda_c/z_c$. We
also analyze the fluctuation-dissipation relationship at the
critical temperature, obtaining the critical
fluctuation-dissipation ratio $X_\infty$. By comparing our results
with the aging scaling reported previously for a model of
interacting flux lines in the vortex glass regime, we found that
the exponents for both models are very different.

\end{abstract}


\maketitle

%.....................................................................
\section{\label{S1}INTRODUCTION}

The possibility of a vortex glass phase in high-$T_c$
superconductors was proposed in 1989 by M. Fisher and coworkers.
\cite{Fisher,review1,review2}
Throughout the years,  two possible scenarios have
been discussed: either there is a continuous phase transition from
the vortex liquid to the vortex glass at a finite critical
temperature,\cite{Fisher,Huse,Reger,koch,petrean} or there is a crossover
temperature\cite{Bokil,reich2000,strachan} below which the
vortex liquid freezes in a glassy regime. While early theoretical\cite{Huse,Reger}
and experimental studies\cite{koch} supported the existence of a finite
temperature critical point,
more recent theoretical arguments tend to favor
the freezing scenario for the vortex liquid at a  crossover
temperature.\cite{Bokil,reich2000} Experimentally, it is difficult
to differentiate between a critical point and a strong crossover
in  near equilibrium measurements in superconductors
with random pinning.\cite{petrean,strachan}
%However, from the experimental point of view it seems
%difficult to resolve among the two possibilities.
The aim of this
work is to show that the study of the {\it nonequilibrium}
dynamics can help to clearly distinguish among these two scenarios. To this
end, we will study nonequilibrium correlation and response
functions in a model for the vortex glass transition that has a
finite temperature critical point (the unscreened gauge glass
model)\cite{Huse} and compare with previous results in a model that has a
crossover freezing transition to a glassy regime (the interacting
flux lines model).\cite{reich2000}

The gauge glass (GG) model\cite{Huse,Reger,moore}  is  a paradigmatic
model for the vortex glass phase transition.\cite{review1,review2}
It consists on the XY model with a random gauge potential vector.
The GG is assumed to represent the physics of a superconductor
with disorder at relatively high magnetic fields.
\cite{review1,review2,Huse} It has been found that this model has
a finite $T_c$ in three dimensions \cite{Reger,gingras,cieplack}
and several numerical works have studied its critical behavior,
considering both equilibrium properties as well as dynamical and
transport properties.
\cite{Huse,Reger,moore,gingras,cieplack,grempel,Olson1,Katzgraber0,Katzgraber1,Katzgraber2}
 Early experiments \cite{koch} on
the scaling of current-voltage curves in high-$T_c$
superconductors obtained critical exponents $\nu$ and $z$ in
reasonable agreement with the exponents obtained in simulations in
the GG. However, in 1995 Bokil and Young\cite{Bokil} showed that
$T_c=0$ in three dimensions when  magnetic screening in a finite
length scale $\lambda$ is included in the GG (by adding
fluctuations of the vector potential).
\cite{Bokil,WY96,WY97,rieger} Later experiments also showed that
the current-voltage curves do not scale as
expected,\cite{strachan} while other experimental groups report
good scaling with the GG exponents.\cite{petrean} Since all
superconductors have a finite screening length $\lambda$ (the
London penetration depth), the findings of Ref.~\onlinecite{Bokil}
discard the scenario of a finite temperature critical point.
Taking into account this result,
Zimanyi and coworkers\cite{reich2000,zim98} considered a model of
interacting flux lines (IFL) with finite interaction length
$\lambda$ and in the presence of quenched disorder.\cite{wallin}
They found in this model that upon decreasing temperature the
vortex-glass-like criticality is arrested  at a crossover
temperature signaling a freezing of flux lines. Below this
temperature the time scales grow very quickly with a dynamical
behavior similar to what is
found in structural (``window'') glasses.\cite{reich2000,Busting}

This later result brings into attention the importance of the
nonequilibrium behavior of vortex glasses as well as its possible
relationship with the dynamical behavior of other glassy systems.
In the recent years,
there has been some  progress in the general
understanding of glasses through the study of their out-of-equilibrium dynamics.\cite{leshouches}
A characteristic of relaxing glassy systems is the loss of stationarity
reflected by their \textit{aging} properties, meaning that the
dynamics of the system depends on the time elapsed after the preparation of
the sample, $t_w$.\cite{leshouches}  As a consequence, dynamic
correlation functions dependend on two times, the ``waiting
time'', $t_w$, and the time $t$ elapsed during the measurement.  Also
the linear response functions  show aging effects,
being dependent on $t_w$ and $t$, and they are anomalous in the sense
that they are not related to their associated correlation functions by
the equilibrium fluctuation-dissipation theorem.~\cite{leshouches,crisanti}
 The standard protocol for the study of aging in glassy
systems\cite{leshouches,crisanti}
consists on preparing the sample at a high temperature
($T_{start}\rightarrow \infty$) and then to quench it to a final
low temperature $T_f$. At $T_f$,
 two-time correlation functions $C(t,t_w)$ and response functions
$R(t,t_w)$ are analyzed, where the measurement starts for $t \ge
t_w$. In several systems a simple aging law with $C(t,t_w)\sim
C_{ag}(t/t_w)$ is typically found. \cite{crisanti} On the other
hand, a ``multiplicative'' aging law of the form $C(t,t_w)\sim
t_w^{-\alpha}C_{ag}(t/t_w)$ has been found in polymers in random
media \cite{yoshino} and in systems that are quenched at the
critical temperature, like ferromagnetic spin
models\cite{Calabrese1,Godreche1,Godreche2,Berthier1,Abriet1,Abriet2}
and the Ising sping glass.\cite{Henkel1,Pleimling1}

In the case of vortices in superconductors with quenched disorder,
Bustingorry {\it et al.}\cite{Busting}  have studied the
nonequilibrium aging dynamics of the IFL model. They have found
that it can  be characterized by ``multiplicative aging'', with a
scaling form similar to what is found in polymers in random
media.\cite{yoshino} In order to compare with this result, we will
study here the the aging and nonequilibrium dynamics of the GG
model at the critical temperature, calculating two-times
correlation and response functions. Being at the critical point, a
multiplicative aging law is expected, which in form should be
similar to the type of aging observed numerically in the
IFL.\cite{Busting} Our aim is to compare the aging exponents of
the GG at $T_c$ with the ones obtained for the IFL in
Ref.~\onlinecite{Busting}. As we will show here, they turn out to
be very different, a result that indicates that nonequilibrium
aging experimental measurements could clearly distinguish among
the two scenarios discussed in the first paragraph.

The paper is organized as follows. In Section II we present the
model Hamiltonian and the simulation method. In Section III we
define the observables to be calculated. In Section IV we present
our results for the correlation and response functions after a
critical quench in the GG and we analyze their scaling with $t_w$.
Finally in Section V we discuss our results comparing with the
disorder-free 3D XY model, the 3D ISG and the IFL.

%.....................................................................
\section{\label{S2} MODEL AND MONTE CARLO SIMULATIONS}

The hamiltonian of the three-dimensional (3D) GG model \cite{Huse}
is given by
\begin{equation}
H = -J \sum_{( i,j )} \cos(\theta_{i} - \theta_{j} - A_{i,j}),
\label{hamiltonian}
\end{equation}
where $\theta_{i}$ represents
the superconducting phase at site $i$, and we
sum over nearest neighbors $( i,j )$ on a cubic
lattice of linear size $L$ ($N=L^3$). The $A_{i,j}$ are
quenched random variables uniformly distributed in the $[0,2\pi]$
interval ($A_{i,j}=-A_{j,i}$); and $J$ is the
coupling between nearest neighbors.
The phases $\theta_i$ can be represented as the
angle of classical two-dimensional spins of
unit length, $\bm{S}_i =(\cos\theta_i,\sin\theta_i)$.
In this work periodic boundary conditions are applied in the phases $\theta_i$.
The energy is normalized in units of $J$, temperature in units of $J/k_B$,
and time scales are measured in number of Monte Carlo steps
(for full sweep for the $N$ sites).

The out-of-equilibrium protocol used in this work, consists on a
quench at time $t=0$ from a $T=\infty$ state to the critical
temperature $T_c$. For the 3D GG model, the critical temperature
is $T_c = 0.46(1)$, according to Ref.~\onlinecite{Katzgraber1}.
From this initial condition different two-times quantities are
analyzed, which depend on both the waiting time $t_w$, when the
measurement begins, and a given time $t>t_w$.

For the Monte Carlo simulation local changes in the phases
$\theta_{i} \to \theta'_{i}$ are accepted with probability given
by the Metropolis rate
\begin{equation}
p( \theta_{i} \to \theta'_{i} ) =\min \lbrace 1, \exp(-\beta
\Delta H) \rbrace. \label{rate}
\end{equation}
Here $\beta$ is the inverse temperature and $\Delta H$ is the
energy difference corresponding to the proposed phase change. In
equilibrium simulations, the acceptance window for $\theta'_{i}$
is usually chosen less than $2\pi$ and dependent on temperature in
order to optimize the updating procedure. Because we are
interested in studying a nonequilibrium process, for local phase
changes we will use the full $2\pi$ acceptance angle window for
the new phases $\theta'_{i}$, following the criterion used in
Ref.~\onlinecite{Katzgraber1,Katzgraber2}. This is done in order
to avoid the possibility that a limited acceptance angle might
introduce an artificial temperature dependence in the relaxation.
%.....................................................................
\section{\label{S3} OBSERVABLES}

For simplicity, we will focus on the study of the two-times autocorrelation function
defined as
\begin{equation}
C(t,t_w) = \frac{1}{N} {\left[ \left \langle \sum_{i=1}^N \cos
\lbrack \theta_{i}(t)-\theta_{i}(t_w) \rbrack \right \rangle
\right]}_{av}, \label{correlation}
\end{equation}
where $\langle ... \rangle$ indicates an average over different
thermal histories (different initial configurations and
realizations of the thermal noise), and $[...]_{av}$ represents a
disorder average over different samples (different realizations of
$A_{i,j}$).
% To calculate this quantity it is only  necessary to store
%the system configuration at $t_w$.

The corresponding two-times linear autoresponse function is
\begin{equation}
R(t,t_w) = \frac{1}{N} {\left[ \left \langle \sum_{i=1}^N
\frac{\delta \bm S_i(t) }{\delta \bm h_i(t_w)} \right \rangle
\right]}_{av}. \label{response}
\end{equation}
In this case, the system is perturbed by applying an infinitesimal
external field $\bm h_i$ conjugated to $\bm S_i$. This corresponds
to a perturbing term in the hamiltonian
$\Delta H = - \sum_i {\bm h_i}\cdot{\bm S_i}$.
In numerical simulations it is more convenient to calculate
the integrated responses: either the thermoremanent response
$M_{TRM}(t,t_w)$ or the zero-field-cooled response
$M_{ZFC}(t,t_w)$, which are obtained by switching on the perturbation
only for times $t < t_w$ and $t > t_w$, respectively.
If we define the reduced integrated responses by
\begin{eqnarray}
\rho_{TRM}(t,t_w) = \frac{T}{h}M_{TRM}(t,t_w) \\
\rho_{ZFC}(t,t_w) = \frac{T}{h}M_{ZFC}(t,t_w),
\end{eqnarray}
then we can relate these functions to $R(t,t_w)$:
\cite{Godreche1}
\begin{eqnarray}
\rho_{TRM}(t,t_w) = T \int_0^{t_w} du R(t,u) \label{integral}   \\
\rho_{ZFC}(t,t_w) = T \int_{t_w}^t du R(t,u).
\end{eqnarray}

 If the perturbing field  in each site, $\bm h_i$, is random and its two
components are independently drawn from a bimodal distribution
$\pm h$, both reduced integrated responses are given by
\cite{Berthier1}
\begin{equation}
\rho(t,t_w) = \frac{T}{h^2 N} {\left[ \left \langle \sum_{i=1}^N
\bm h_i \cdot \bm S_i \right \rangle \right]}_{av}.
\label{response2}
\end{equation}

In this work we show simulation results for systems of size $L = 30$ at the
temperature $T_c = 0.46$. We take the disorder average over
60 samples and, for each sample, we carry out a number of
$10$ thermal histories. The magnitude of the external
fields used for the calculation of the response function
was  $h =0.1$ and $h = 0.05$.

%.....................................................................
\section{\label{S4}CRITICAL QUENCH}

At the critical temperature  $T_c$ the equilibrium autocorrelation
relaxation time $\tau_c$ increases with the size as $\tau_c \sim
L^{z_c}$, where $z_c$ is the dynamical critical exponent. For
simple ferromagnetic systems, it is known that in a critical
quench spatial correlations over a length scale of $l \sim
t^{1/z_c}$ are just as in the equilibrium critical state. This
means that the system appears critical on scales smaller than $l$,
while it appears  disordered on larger scales. \cite{Godreche1}

\begin{figure}[t]
\includegraphics[width=7.5cm,clip=true]{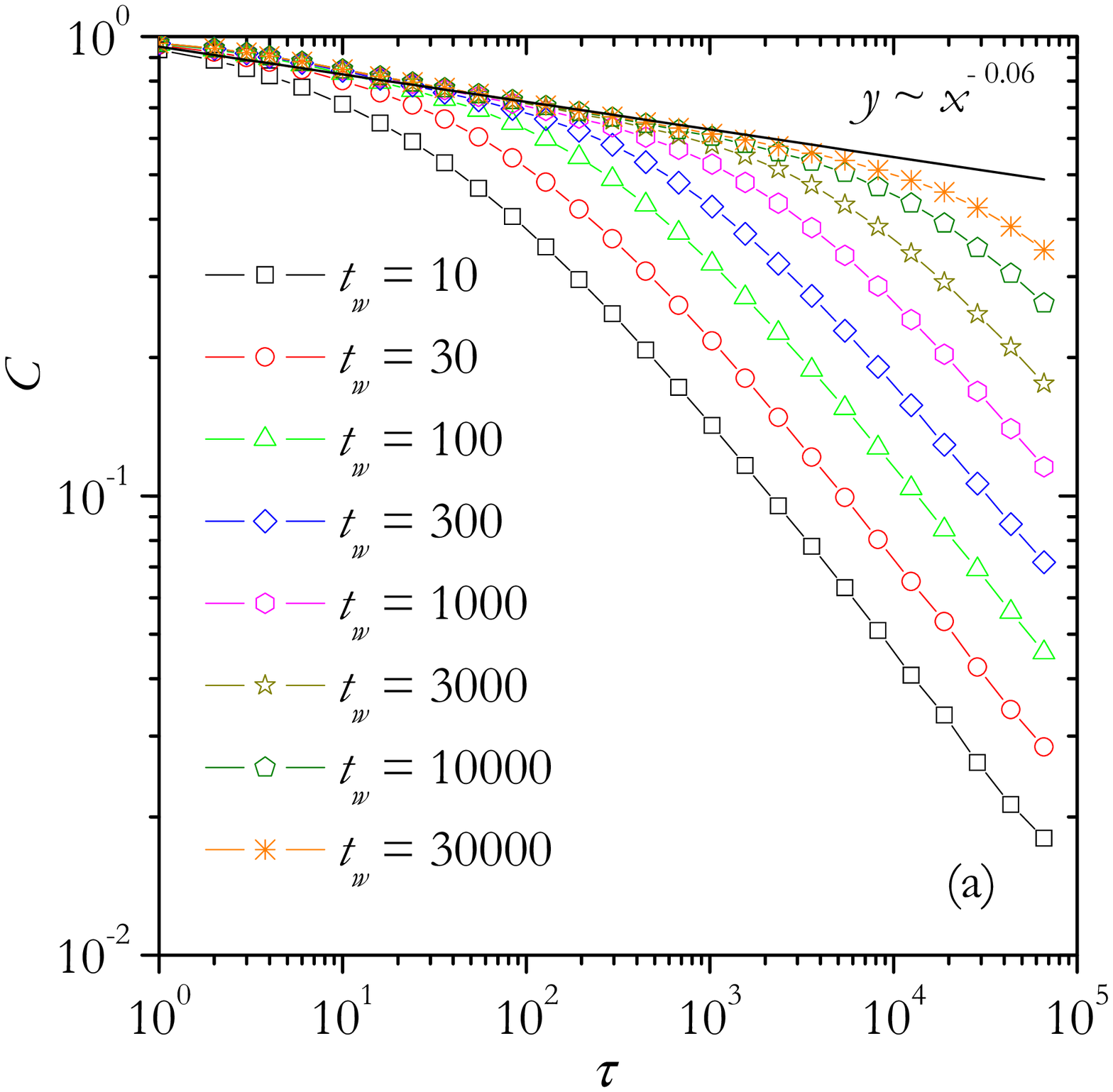}
\includegraphics[width=7.5cm,clip=true]{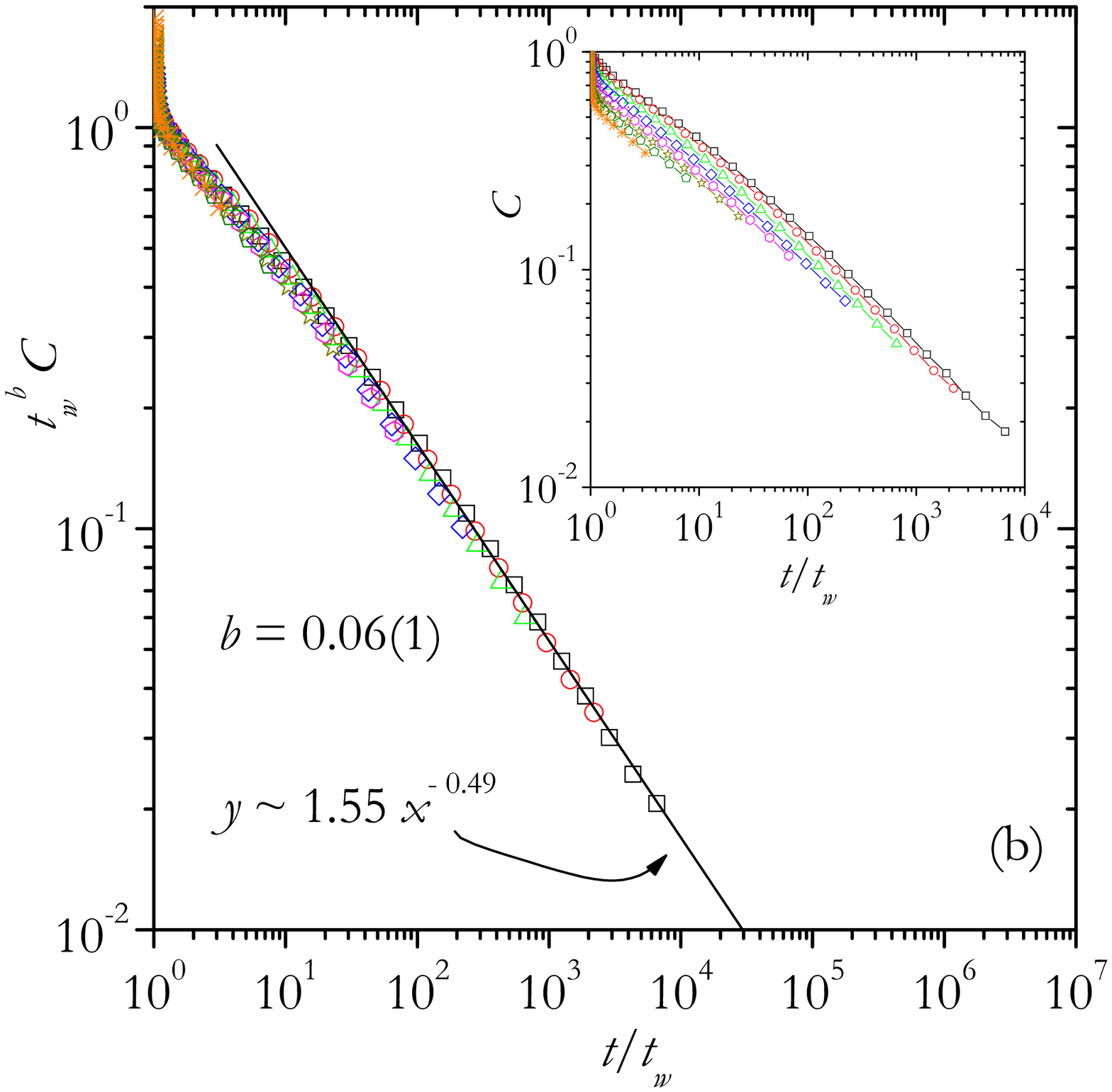}
\caption{\label{figure1} (Color online) (a) Autocorrelation
function $C$ vs $\tau$ for eight different waiting times as
indicated.  (b) Data collapsing.  Inset: autocorrelation function
$C$ vs $t/t_w$.  Time scales are measured in number of Monte Carlo steps
(for full sweep for the $N$ sites). }
\end{figure}

Considering this, the autocorrelation function (\ref{correlation}) is expected to
behave as \cite{Godreche1,Pleimling1}
\begin{equation}
C(t,t_w) = {t_w}^{-b} f_C \left ( \frac{t}{t_w} \right ).
\label{co2}
\end{equation}
For $\tau \ll t_w $ ($\tau = t - t_w$), the critical scaling
function $f_C (x)$ behaves as
\begin{equation}
f_C (x) \sim [x-1]^{-b}.
\end{equation}
For $d$ dimensional spin glasses
it has been found\cite{Ogielski1,Henkel1} that
\begin{equation}
b = \frac{d-2+\eta}{2 z_c}. \label{bexponent}
\end{equation}
Here $\eta$ is the static critical exponent associated to the pair
correlation function.
Then, for $\tau \ll t_w $ the autocorrelation function is
\begin{equation}
C(t,t_w) \sim \tau^{-b}.
\end{equation}
On the other hand, when $\tau \gg t_w $ we have:
\begin{equation}
f_C (x) \sim A_C x^{-\lambda_C / z_c}, \label{co3}
\end{equation}
where $A_C$ is a constant and $\lambda_C$ is known as the
autocorrelation exponent. \cite{Henkel1}

The Fig.~\ref{figure1}(a) shows the behavior of the
autocorrelation function $C(t,t_w)$ vs $\tau$ for eight different
waiting times $t_w$. For $\tau \ll t_w $ we can fit a power law
behavior, $\tau^{-b}$, with exponent $b \approx 0.06$, as shown in
the plot. For $\tau > t_w$ we observe that the correlation
function shows aging since it depends strongly on the waiting time
$t_w$. In the inset of Fig.~\ref{figure1} (b) we show the result
of assuming ``simple aging'' by plotting $C$ as a function of the
ratio $t/t_w$. We observe that the data does not show a good
collapse in a single curve in this case. A good data collapse can
be obtained if the ``multiplicative aging'' scaling form of
Eq.(\ref{co2}) is assumed. In  Fig.~\ref{figure1} (b), we show the
plot of $t_w^bC$ vs. $t/t_w$, obtaining good scaling with $b =
0.06(1)$. This value is in good  agreement with
Eq.~(\ref{bexponent}): if we take from previous simulations of the
3D GG model the values $\eta = -0.47(2)$ \cite{Katzgraber1} and
$z_c = 4.5(1)$ \cite{Katzgraber2} we obtain $b = 0.059(4)$.  We
now consider the case of $\tau\gg t_w$. In the Fig.~\ref{figure1}
(b) we show that in this case the asymptotic behavior of $C$ has a
power law form consistent with Eq.(\ref{co3}). We obtain $A_C =
1.55(5)$ and $\lambda_c / z_c = 0.49(2)$.

We now consider the autoresponse function. In this case the
scaling equation is \cite{Godreche1}
\begin{equation}
R(t,t_w) = {t_w}^{-1-a} f_R \left ( \frac{t}{t_w} \right ),
\label{resp2}
\end{equation}
For $\tau \gg t_w $ is expected that
$$f_R (x) \sim A_R x^{-\lambda_R / z_c}$$
where $A_R$ is a constant amplitude. Then, if
we consider Eq.~(\ref{integral}), we obtain the thermoremanent
reduced integrated response
\begin{equation}
\rho_{TRM}(t,t_w) = {t_w}^{-a} f_\rho \left ( \frac{t}{t_w} \right
), \label{re2}
\end{equation}
where
\begin{equation}
f_\rho (x) \sim A_\rho x^{-\lambda_R / z_c}, \label{re3}
\end{equation}
and $A_\rho$ is a constant.  For critical systems it is
expected that $a=b$, and $\lambda_C=\lambda_R$.

\begin{figure}[t]
\includegraphics[width=7.5cm,clip=true]{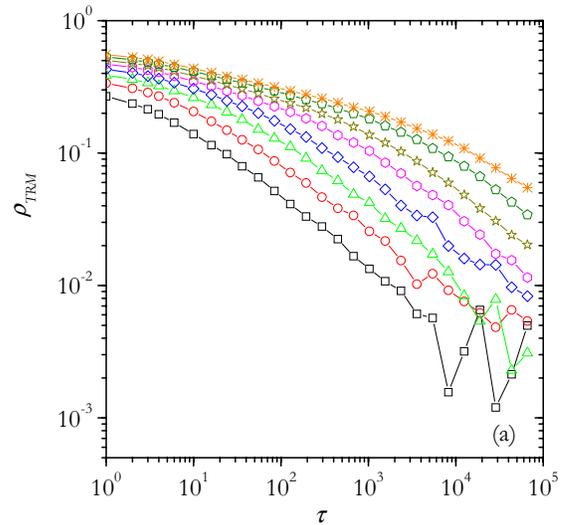}
\includegraphics[width=7.5cm,clip=true]{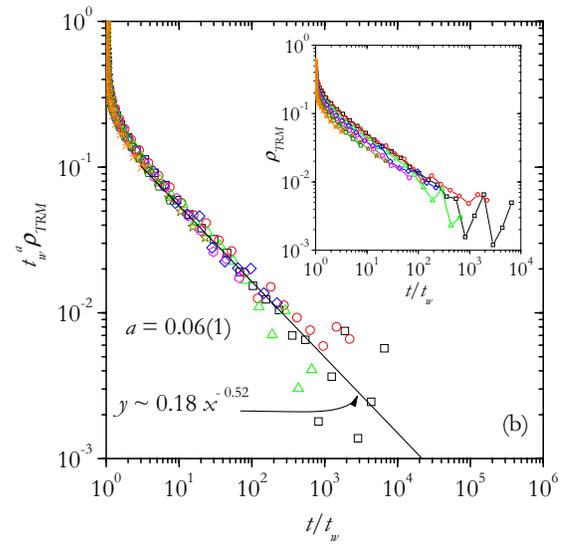}
\caption{\label{figure2} (Color online) (a) Function $\rho_{TRM}$
vs $\tau$ for different waiting times and $h=0.05$.  All symbols
are the same as in Fig.~\ref{figure1} (a).  (b) Data collapsing.
Inset: Function $\rho_{TRM}$ vs $t/t_w$ for different waiting
times and $h=0.05$.}
\end{figure}

In Fig.~\ref{figure2} (a) we show the behavior of  $\rho_{TRM}$
function vs $\tau$ for different waiting times and $h=0.05$,
observing that the response function also shows aging. In the
inset of Fig.~\ref{figure2} (b) we show $\rho_{TRM}$ vs $t/t_w$:
no data collapse is observed. In Fig.~\ref{figure2} (b) we plot
$t_w^a\rho_{TRM}$ vs $t/t_w$, obtaining good data collapse for $a
= 0.06(1)$. The asymptotic behavior observed for large temporal
separations ($\tau\gg t_w$) can be fitted with
 a power law with $A_\rho= 0.18(2)$ and $\lambda_R / z_c = 0.52(2)$.

At thermodynamic equilibrium the fluctuation-dissipation
theorem (FDT) is satisfied, and therefore it is possible to draw a simple
expression to relate the correlation with the thermoremanent
reduced integrated response
\begin{equation}
\rho_{TRM}(t-t_w) = C(t-t_w).
\end{equation}
Then, for a parametric plot of $\rho_{TRM}$ vs. $C$, we should obtain a
straight line of slope $+1$ in equilibrium.
On the other hand, for a nonequilibrium process the FDT is not fulfilled,
and it has been proposed a generalized relation of the form
\begin{equation}
\rho_{TRM}(t,t_w) = X(t,t_w) C(t,t_w), \label{QFDT}
\end{equation}
where $X(t,t_w)$ is called the fluctuation-dissipation ratio
(FDR). \cite{Cugliandolo} For $1 \ll t_w \ll t$ we can estimate
the limit :
$$X_\infty = \lim_{t_w\rightarrow\infty} \lim_{t\rightarrow\infty} X(t,t_w).$$
For a critical quench, $X_\infty$ is
expected to be universal in the sense that it does not depend neither on the
initial conditions nor on the details of the dynamics.
\cite{Godreche2}

\begin{figure}
\includegraphics[width=7.5cm,clip=true]{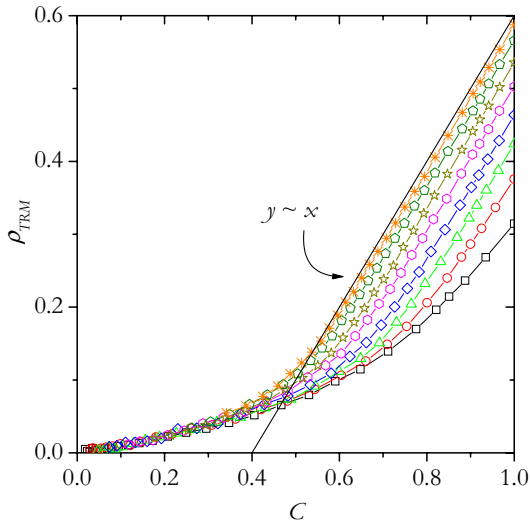}
\caption{\label{figure3} (Color online) FDT plot for different
waiting times and $h=0.05$.  All symbols are the same as in
Fig.~\ref{figure1} (a).}
\end{figure}

The Fig.~\ref{figure3} shows the  $t$-parametric plot of $\rho_{TRM}(t,t_w)$ vs $C(t,t_w)$
for different waiting times $t_w$.  In the quasi-equilibrium regime ($\tau \ll t_w$) the
FDT holds and $X=1$. This is observed in Fig.~\ref{figure3} where
for large $t_w$ the slope in the FDT plot tends to $1$.  On the
other hand, when $1 \ll t_w \ll t$, we observe that $X < 1$ and
FDT is violated.
From Eqs.~(\ref{co2}), (\ref{co3}), (\ref{re2}), (\ref{re3}) and
(\ref{QFDT}), it is possible to demonstrate that
$$X_\infty = A_\rho /A_C. $$
From the amplitudes $A_C$ and $A_\rho$ calculated previously
we obtain $X_\infty = 0.12(2)$.

\begin{figure}
\includegraphics[width=7.5cm,clip=true]{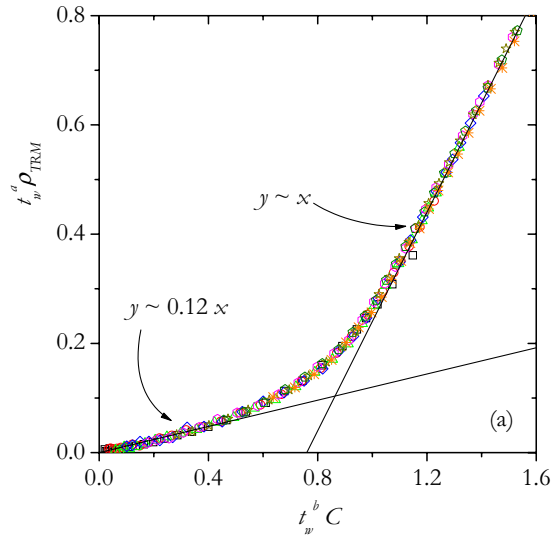}
\includegraphics[width=7.5cm,clip=true]{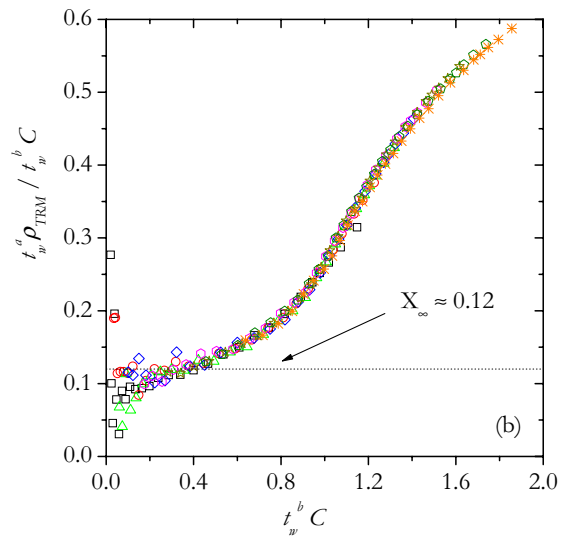}
\caption{\label{figure4} (Color online) (a) Data collapsing for
all curves in Fig.~\ref{figure3}. (b) Plot of ratio ${t_w}^a
\rho_{TRM}/{t_w}^b C$ vs ${t_w}^b C$. All symbols are the same as
in Fig.~\ref{figure1} (a).}
\end{figure}

We can obtain $X_\infty$ numerically from the FDT plot of Figure 3 in two
different ways.
First, in  Fig.~\ref{figure4} (a) we show a data
collapse  plotting $\rho_{TRM}(t,t_w) t_w^a$ vs $C(t,t_w)t_w^b$.
The slope near the origin corresponds to the limit of FDR, obtaining $X_\infty=0.12(1)$.
Second, in Fig.~\ref{figure4} (b) we plot the ratio
${t_w}^a \rho_{TRM}/{t_w}^b C$ vs ${t_w}^b C$ for all times $t$
and $t_w$. We see that the ordinate at the origin is compatible with
$X_\infty\approx 0.12$.

%.....................................................................
\section{\label{S5}DISCUSSION}

We have therefore obtained the exponents that characterize the
aging scaling of the 3D GG model at the critical temperature,
obtaining $b=0.06(1)$, $\lambda_C/z_c=0.49(2)$, $a=0.06(1)$,
$\lambda_R/z_c=0.52(2)$, and $X_\infty=0.12(1)$. A comparison with
results in the corresponding model without disorder, {\it i.e.}
the 3D XY model, shows that the exponents are very different, as
expected. Abriet and Karevski,\cite{Abriet2} have found $a=b =
1/2$, $\lambda_C/z=\lambda_R/z=1.34$, which shows that in the
disorder-free case the ``multiplicative'' scaling of aging is stronger
(i.e., much larger values of $a$ and $b$). Also a very different
limit of the FDR,  $X_\infty=0.43(4)$, was obtained, as expected
since they should belong to different universality classes.

On the other hand, the critical aging of the GG turns out to be
very similar to results found in the 3D ISG.
\cite{Henkel1,Pleimling1} It  has been found for a binary  distribution
in the exchange couplings:
$a=0.060(4)$, $b =0.056(3)$, $\lambda_C/z=0.362(5)$,
$\lambda_R/z=0.38(2)$ and $X_\infty=0.12(1)$; and for a Gaussian
distribution in the exchange couplings:
$a=0.044(1)$, $b =0.043(1)$, $\lambda_C/z=0.320(5)$,
$\lambda_R/z=0.33(2)$ and $X_\infty=0.09(1)$. \cite{Henkel1} As we
can see, the aging exponents are slightly smaller in the 3D ISG
(meaning a weaker multiplicative aging), but the values of
$X_\infty$ are comparable within the statistical error with the
GG. It has been argued in Ref.~\onlinecite{Calabrese1,Godreche1,Godreche2}
that $X_\infty$ is a novel  universal
quantity of non-equilibrium critical dynamics. It is interesting
to observe that spin glass models like the ISG and
GG have similar values of $X_\infty$.
However, let us remind that most work on the {\it equilibrium} properties of
the GG and the ISG show evidence that they are in different universality
classes.\cite{Reger,gingras,moore,Olson1,Katzgraber2}

As discussed in the introduction, two types of behaviors are
predicted for the vortex glass transition: either a finite
temperature critical point as described by the GG, or a crossover
temperature as found in the IFL. The work of Bokil and
Young\cite{Bokil} discards the first scenario. However, from the
experimental point of view, the issue is not clearly resolved
since several experiments show reasonably good scaling at a
transition temperature.\cite{petrean} Comparing our results with
the behavior found by Bustingorry {\it et al.}\cite{Busting} in
the IFL, we believe that a study of the non-equilibrium dynamics
of the vortex glass transition will be able to clearly distinguish
between the two models. In simulations of the IFL ,
off-equilibrium correlation and response functions are found to
scale as $B(t,t_w) \sim t_w^\alpha \tilde{B}(t/t_w)$, and also it
is found that $\tilde{B}(x) \sim x^\alpha$ for $x\gg
1$.\cite{Busting} The first thing to notice when comparing with
critical aging in the GG is that, in the IFL, the corresponding
multiplicative aging exponent $b$ and the autocorrelation exponent
$\lambda_C/z$ coincide $b=\lambda_C/z=\alpha$ (and similarly for
the response exponents, $a=\lambda_R/z=\alpha$). The second main
difference is that the ``multiplicative'' aging in the IFL is very
strong, $b_{\rm IFL}=\alpha \approx 0.25-0.35$  when compared with
the GG at the critical point, $b_{\rm GG} \approx 0.06$.
In the IFL the competition between the elasticity of each
flux line and the randomness of the pinning potential leads to
a very slow dynamics and strong aging with a large $\alpha$ exponent.
Moreover, $\alpha$ depends with the disorder strength in the
IFL, being larger for stronger disorder.

In this work we have analyzed the correlation and integrated
response functions defined in Eqs.(\ref{correlation}) and
(\ref{response2}), since they are easy to compute numerically. In
experiments in superconductors, correlation (response) functions
of other observables like magnetization (magnetic susceptibility)
and voltage (resistance) are, of course, more available. Aging
scaling similar to Eqs.(\ref{co2}) and (\ref{re2}) will also be
expected in these other physical quantities. For example, one
possible way of observing experimentally the out-of-equilibrium
dynamics of the vortex glass is monitoring the magnetic relaxation
as done in Ref.\onlinecite{papa} for granular samples. Another
possibility is to perform electrical transport experiments near
the transition temperature following a protocol similar to the one
used by Ovadyahu and coworkers to study the electron glass regime
in Anderson insulators.~\cite{ovadyahu} In either case, an
analysis of the multiplicative aging exponent will clearly
distinguish among the two different models ($b_{\rm IFL} \approx
0.25-0.35$ or $b_{\rm GG} \approx 0.06$).

%.....................................................................
\begin{acknowledgments}
We acknowledge support from CNEA (Argentina) and from CONICET (Argentina)
under project PIP5596.
FR thanks Universidad Nacional de San Luis (Argentina) under
project 322000 and support from CONICET
under project PIP6294.
DD thanks support from ANPCyT (Argentina)
under projects PICT2003-13829 and PICT2003-13511.
\end{acknowledgments}
%.....................................................................

%.....................................................................

\begin{thebibliography}{99}

%%% vortex  glass

\bibitem{Fisher}
M. P. A. Fisher, Phys. Rev. Lett. {\bf 62}, 1415 (1989); D. S.
Fisher, M. P. A. Fisher, and D. A. Huse, Phys. Rev. B {\bf 43},
130 (1991).

%%% reviews vortex glass

\bibitem{review1} G. Blatter, M. V. Feigel'man, V. B. Geshkenbein, A. I. Larkin, and V. M. Vinokur,
Rev. Mod. Phys. {\bf 66}, 1125 (1994).

\bibitem{review2} T. Nattermann and S. Scheidl,
Adv. in Phys. {\bf 49}, 607 (2000).

%% gauge glasss

%\bibitem{Shih1} W. Y. Shih, C. Ebner, and D. Stroud, Phys. Rev. B {\bf 30}, 134 (1984).

\bibitem{Huse}%defines GG first time, some critical properties
D. A. Huse and H. S. Seung, Phys. Rev. B {\bf 42}, R1059 (1990).

\bibitem{Reger}% transition in d=3, critical exponents at Tc and T=0
J. D. Reger, T. A. Tokuyasu, A. P. Young, and M. P. A. Fisher,
Phys. Rev. B {\bf 44}, 7147 (1991).

%%%experiments vortex glass

\bibitem{koch}
R. H. Koch, V. Foglietti, W. J. Gallagher, G. Koren, A. Gupta, and
M. P. A. Fisher, Phys. Rev. Lett. {\bf 63}, 1511 (1989).

\bibitem{petrean} A. M. Petrean, L. M. Paulius, W.-K. Kwok, J. A. Fendrich, and G. W. Crabtree,
Phys. Rev. Lett. {\bf 84}, 5852 (2000).


%screening and ifl
\bibitem{Bokil} H. S. Bokil and A. P. Young, Phys. Rev. Lett. {\bf 74},
3021 (1995).

\bibitem{reich2000} C. Reichhardt, A. van Otterlo, and G. T. Zim\'anyi,
Phys. Rev. Lett. {\bf 84}, 1994 (2000).

\bibitem{strachan} D. R. Strachan, M. C. Sullivan, P. Fournier, S. P. Pai, T. Venkatesan, and C. J. Lobb,
Phys. Rev. Lett. {\bf 87}, 067007 (2001);
I. L. Landau and H. R. Ott, Phys. Rev. B {\bf 65}, 064511 (2002).


%%more gaugle glass
\bibitem{moore}A. Houghton and M. A. Moore,
 Phys. Rev. B {\bf 38}, 5045 (1988);
 M. A. Moore and S. Murphy,
 Phys. Rev. B {\bf 42}, 2587 (1990).

\bibitem{gingras}
M. J. P. Gingras, Phys. Rev. B {\bf 44}, 7139 (1991); Phys. Rev. B
{\bf 45}, 7547 (1992)

\bibitem{cieplack} M. Cieplak,
J. R. Banavar, and A. Khurana, J. Phys. A {\bf 24}, L145 (1991).
%possibility of transition in d=3.

\bibitem{grempel} J. Maucourt and D. R. Grempel, Phys. Rev. B {\bf 58}, 2654 (1998)
%transition in d=3, domain walls at T=0

\bibitem{Olson1} T. Olson and A. P. Young, Phys. Rev. B {\bf 61}, 12467 (2000).
%transtion in d=3, Tc, critical exponents, compare with ISG.

\bibitem{Katzgraber0} H. G. Katzgraber and A. P. Young, Phys. Rev. B {\bf 64}, 104426 (2001);
Phys. Rev. B {\bf 66}, 224507 (2002).
% GG d=3 at low T, stiffness exponent

\bibitem{Katzgraber1} H. G. Katzgraber and I. A. Campbell, Phys. Rev. B {\bf 69}, 094413 (2004).
%critical properties and critical exponentes, d=2,3,4. Dynamical properties also.

\bibitem{Katzgraber2} H. G. Katzgraber and I. A. Campbell, Phys. Rev. B {\bf 72}, 014462 (2005).
%critical dynamics, exponent z, for GG and ISG






%more screening effects, absence of transtion


\bibitem{WY96}
C. Wengel and A. P. Young, Phys. Rev. B {\bf 54}, R6869 (1996).
%absence of transition with screening, resistivity, etc.

\bibitem{WY97}
C. Wengel and A. P. Young, Phys. Rev. B {\bf 56}, 5918 (1997).
%with and without screening, resistivity, comparison with XYSG.

\bibitem{rieger} J. Kisker and H. Rieger, Phys. Rev. B {\bf 58},
R8873 (1998).
%T=0 in model with screening, absence of transition


%%%interacting flux lines


\bibitem{zim98} A. van Otterlo, R. T. Scalettar, and G. T. Zim\'anyi,
Phys. Rev. Lett. {\bf 81}, 1497 (1998).

\bibitem{wallin} It is woth mentioning that in simulations of a model of flux lines with
long-range interactions ($\lambda=\infty$) a finite critical temperature was found,
but with critical exponents different from the GG.
See A. Vestergren, J. Lidmar, and M. Wallin, Phys. Rev. Lett. {\bf 88}, 117004 (2002).

\bibitem{Busting} S. Bustingorry, L. F. Cugliandolo, and D. Dom\'{\i}nguez,
       Phys. Rev. Lett. {\bf 96}, 027001 (2006);Phys. Rev. B {\bf 75}, 024506 (2007).

%%%experiments vortex glass







%% glasses

\bibitem{leshouches} L. F. Cugliandolo, in {\it Slow Relaxations and Nonequilibrium Dynamics in Condensed Matter},
ed. by J. -L. Barrat, J. Dalibard, M. Feigel'man, and J. Kurchan
(Springer, Berlin, 2002).

\bibitem{crisanti} A. Crisanti and F. Ritort, J. Phys. A {\bf 36}, 181 (2003).

%%polymers
\bibitem{yoshino} H. Yoshino, Phys. Rev. Lett.
{\bf 81}, 1493 (1998).

%% critical aging

\bibitem{Calabrese1} P. Calabrese and A. Gambassi, J. Phys. A: Math. Gen. {\bf 38},
R133 (2005).
%review of critical aging

\bibitem{Godreche1} C. Godr\`{e}che and J. M. Luck, J. Phys.: Condens. Matter {\bf 14}, 1589 (2002).
%critical aging in ferromagnetic systems, Ising and spherical

\bibitem{Godreche2} C. Godr\`{e}che and J. M. Luck, J. Phys. A: Math. Gen. {\bf 33}, 1151 (2000); J. Phys. A: Math. Gen. {\bf 33}, 9141 (2000).


\bibitem{Berthier1} L. Berthier, P. C. W. Holdsworth, and M. Sellito, J. Phys. A: Math. Gen. {\bf 34}, 1805 (2001).
%aging in 2D XY model


\bibitem{Abriet1} S. Abriet and D. Karevski, Eur. Phys. J. B {\bf 37},
47 (2004).
%aging in 2D XY model

\bibitem{Abriet2} S. Abriet and D. Karevski, Eur. Phys. J. B {\bf 41}, 79
(2004).
%aging in 3D XY

\bibitem{Henkel1} M. Henkel and M. Pleimling, Europhys. Lett. {\bf 69}, 524 (2005).
%critical aging in ISG, d=4,d=4, binary and gaussian


\bibitem{Pleimling1} M. Pleimling and I. A. Campbell, Phys. Rev. B {\bf 72}, 184429 (2005).
%critical aging in ISG


\bibitem{Ogielski1} A. T. Ogielski, Phys. Rev. B {\bf 32}, 7384 (1985).

\bibitem{Cugliandolo} L. F. Cugliandolo and J. Kurchan, J. Phys. A: Math. Gen. {\bf 27}, 5749 (1994).


% experiments on aging
\bibitem{papa}
E. L. Papadopoulou, P. Nordblad, P. Svedlindh, R. Sch\"oneberger,
and R. Gross, Phys. Rev. Lett. {\bf 82}, 173 (1999); E. L.
Papadopoulou, P. Svedlindh, and P. Nordblad, Phys. Rev. B {\bf
65}, 144524 (2002).

\bibitem{ovadyahu}
A. Vaknin, Z. Ovadyahu, and M.Pollak, Phys. Rev. Lett. {\bf 84},
3402 (2000); A. Vaknin, Z. Ovadyahu, and M.Pollak, Phys. Rev. B
{\bf 65}, 134208 (2002).

\end{thebibliography}
\end{document}